\documentclass[prl, twocolumn, 
superscriptaddress, groupedaddress, showpacs,
amsfonts]{revtex4}

\usepackage{graphicx}

\begin{document}

\title{Probing the Structure and Energetics of Dislocation Cores in 
SiGe Alloys through Monte Carlo Simulations\footnote{Accepted for publication in Physical Review Letters.}}

\author{Ioannis N. Remediakis}
\email{remed@physics.uoc.gr}
\affiliation{Department of Physics, University of Crete, P.O. Box
2208, 71003 Heraklion, Crete, Greece}
\author{David E. Jesson}
\affiliation{School of Physics, Monash University, Victoria 3800, Australia}
\author{Pantelis C. Kelires}
\affiliation{Department of Physics, University of Crete, P.O. Box
2208, 71003 Heraklion, Crete, Greece}
\date{\today}

\begin{abstract}

We present a methodology for the investigation of dislocation energetics in
segregated alloys based on Monte Carlo simulations which equilibrate the
topology and composition of the dislocation core and its surroundings. An
environment-dependent partitioning of the system total energy into atomic
contributions allows us to link the atomistic picture to continuum elasticity
theory. The method is applied to extract core energies and radii of 60$^\circ$
glide dislocations in segregated SiGe alloys which are inaccessible by other
methods.

\end{abstract}


\maketitle

Dislocations are the most common extended defects in crystalline solids. A
prototypical example in semiconductor physics is offered by the generation of
dislocations in SiGe alloys grown on Si which relieve the stress induced by
the 4\% lattice mismatch between Si and Ge \cite{hull89}. The presence of
dislocations in SiGe alloys has a strong impact on their mechanical and
electronic properties.  Far from the dislocation, the induced strain field
depends only on macroscopic properties of the material, such as its shear
modulus.  However, near the dislocation line, the local properties of the
material deviate significantly both from those of the ideal solid and the
predictions of continuum elasticity theory. This ``core region'' of the
dislocation, accessible only to atomistic calculations, can have different
structures depending on the material, the applied external strain, and other
parameters \cite{hirthbook}.

Although the core structures and energetics of dislocations in elemental
crystals have been well studied \cite{arias94,blase00}, relatively few
investigations have linked the atomistic picture of dislocated alloy systems
to traditional continuum descriptions.  Blumenau {\it et al.}\ used an
atomistic description to calculate core structures and energies in perfectly
ordered alloys, such as SiC \cite{blumenau02}.  Martinelli {\it et al.}\
performed simulations on SiGe alloys and found that Si-rich and Ge-rich
nanowires are formed near the dislocation core \cite{martinelli04}. However,
despite these important contributions, a well-defined method for extracting
core energies and radii from dislocations in segregated alloys has not yet
been proposed. In addition, the importance of alloy segregation and its
relationship to traditional continuum descriptions including the concept of
core parameters is not fully understood.

In view of the crucial role played by dislocations in many important
applications for SiGe materials, such as bulk alloy layers for quantum wells
\cite{houghton90} or nanoislands for quantum dots \cite{ovidko02,denker05}, it
is highly desirable to understand the core energetics and related properties.
In the present work, we develop a framework to calculate the core energies and
core radii of segregated alloys using atomistic calculations, taking a
60$^\circ$ glide dislocation in a Si$_{0.5}$Ge$_{0.5}$ alloy as a prototype.
This is made possible by an environment-dependent partitioning of the total
energy into atomic contributions, which allows us to locally probe the energy
at the dislocation core, and link the atomistic picture to continuum theories.

Our approach is based on state-of-the-art continuous-space Monte Carlo (MC)
simulations, employing the multi-component empirical potentials of Tersoff
\cite{tersoff89}. This method, capable of calculating the free energy of the
system with great statistical precision, has been applied with success in
similar contexts \cite{kelires89,kelires95-98}. Three types of random moves
are involved in the MC algorithm: atomic displacements, volume changes, and
mutual identity exchanges between atoms of different kinds, leading to
topological and compositional equilibration of the system, as described in
Ref.\ \cite{kelires95-98}. The simulation supercell consists of
40$\times$10$\times$8 6-atom unit cells of the diamond lattice, with lattice
vectors having [$0\bar{1}1$], [$1\bar{1}0$] and [$111$] directions, including
a total of 19200 atoms. We checked the convergence of our results with respect
to the size of the simulation cell by running test simulations with different
cell sizes. Epitaxial SiGe alloys grown on Si(100) are simulated by
constraining them to have the lattice constant of Si in the supercell, and
then allowing relaxation of the lattice vectors only in the [100] direction.

The most commonly observed dislocations in the diamond lattice are those
corresponding to the \{111\} slip system; the lowest energy perfect
dislocation in tetrahedral semiconductors is the so-called 60$^\circ$ glide
dislocation \cite{blumenau02}. Although such dislocations usually dissociate
to Shockley partials, especially in diamond, epitaxially grown alloys are very
likely to hold undissociated 60$^\circ$ glide dislocations, due to the strain
introduced by the the lattice mismatch \cite{marzegalli05}. We therefore
choose to study 60$^\circ$ glide dislocations as characteristic prototypes of
dislocations in SiGe alloys. Two such dislocations with opposite Burgers
vectors are introduced in the cell \cite{ASE}, so that the total displacement
field away from the dislocation cores is zero. The distance between the
dislocation cores is about 80 \AA, ensuring that dislocation motion due to
mutual interaction can be neglected for the temperatures of our
simulations. The Burgers vectors and the lines of the two dislocations lie in
the [$0\bar{1}1$] and [$1\bar{1}0$] direction, respectively.

\begin{figure} 
\begin{center}
\includegraphics[width=0.9\columnwidth]{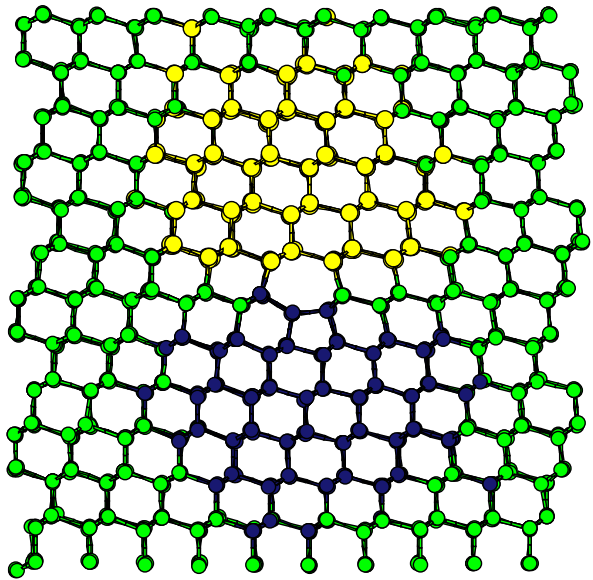} 
\end{center}
\caption{(color online) Equilibrated structure, 2 nm around the core of a
  60$^\circ$ glide dislocation in a Si$_{0.5}$Ge$_{0.5}$ alloy at 300 K. The
  dislocation line is normal to the plane of the plot. Sites with $c\le0.3$,
  having strong preference for Si, are shown in darkest color; sites with
  $c\ge0.7$, having strong preference for Ge, are shown in lightest color. }
\label{fig:structure}
\end{figure}

We begin by equilibrating the unit cell for dislocated bulk SiGe at 300
K. Simulations at higher temperatures (900 K) show similar composition
profiles. After several equilibration steps, including identity exchanges, a
long run ($\sim 10^{10}$ MC steps) follows, over which the average occupancy
of each site, $c$, is calculated. This is defined as the fraction of the
simulation time (MC steps) when a site is occupied by a Ge atom over the whole
run. Its values range between 0 (site always occupied by Si atoms) and 1 (site
always occupied by Ge atoms).  For the bulk Si$_{0.5}$Ge$_{0.5}$ alloy at 300
K, the calculated values of $c$ are $0.5\pm0.02$. However, for the cell
containing a dislocation, there are sites near the dislocation core with
values of $c$ very close to 0.0 and 1.0, showing strong preference for
occupancy by Si or Ge atoms, respectively. This is illustrated in Fig.\
\ref{fig:structure}, which shows the equilibrated atomistic structure of the
dislocation core. It can be observed that the alloy segregates into Si-rich
and Ge-rich cylindrical regions on opposite sides of the dislocation line,
which can be viewed as self-assembled nanowires in the alloy
\cite{martinelli04}. 

\begin{figure} 
\begin{center}
\includegraphics[width=0.99\columnwidth]{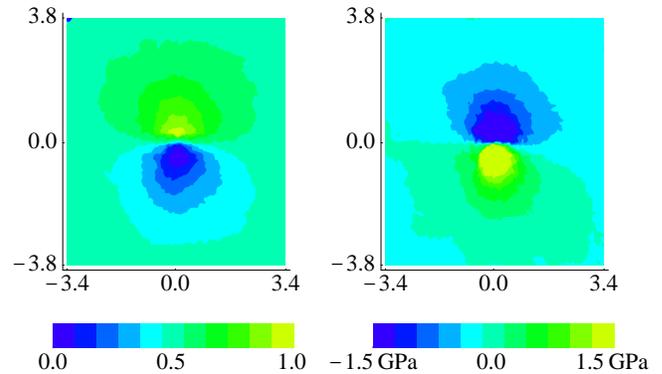} 
\end{center}
\caption{(color online) Contour plots of the average composition (left) and
  average atomic stresses (right) around a 60$^\circ$ glide dislocation in a
  Si$_{0.5}$Ge$_{0.5}$ alloy at 300 K. The dislocation line is normal to the
  plane of the plot. Distances shown on the axes of the plots are given in
  nm.}
\label{fig:compos}
\end{figure}

To explain this effect we note the similarity to stress induced segregation at
the (100) reconstructed surface \cite{kelires89}. The significant near-surface
stress field, resulting from surface dimerization, drives smaller atomic
volume Si atoms to occupy sites under compression, while the Ge atoms prefer
sites under tension. In the present case, this correlation between local
stress and composition can be probed via the atomic stress, defined as
$\sigma_i= - \mathrm{d}E_i/\mathrm{d}{V_i}$, where $E_i$ is the energy of atom
$i$ and $V_i$ is the volume available to atom $i$ \cite{kelires89}. In order
to link the atomistic picture with macroscopic theories of elasticity, we
define a continuous stress field $\sigma(\mathbf{r})=\sum_{i} \sigma_i$. The
sum runs over all atoms $i$ that have Cartesian positions $\mathbf{r}_i$ such
that $|\mathbf{r}_i-\mathbf{r}|<3.0$\ \AA. In the same manner, we define a
continuous composition field. Fig.\ \ref{fig:compos} shows contour plots of
the composition and stress fields on a plane perpendicular to the dislocation
line. There is a clear correlation between the two, as Ge atoms prefer to
occupy sites under tension while Si atoms prefer sites under compression. The
observed composition contours are very similar to those found by lattice Monte
Carlo simulations of dislocations in alloys \cite{wang00}.
$\sigma(\mathbf{r})$ has the same qualitative features as the hydrostatic
component of the stress tensor around a dislocation,
$\sigma(x,y)\equiv(\sigma_{xx}+\sigma_{yy}+\sigma_{zz})/3\sim
-\frac{y}{x^2+y^2}$, as calculated in linear elasticity theory
\cite{hirthbook}. Similar segregation results are found for both relaxed and
epitaxial (100)-constrained alloys.

The significant segregation evident in Fig.\ \ref{fig:structure} will clearly
influence the dislocation core energy and elastic
properties. It is therefore important to develop a framework for defining
dislocation core parameters in the presence of segregation. Our approach is
based on a partitioning of the dislocation formation energy into `elastic' and
`segregation' components. Consider first the radial dislocation formation
energy,
\begin{equation}
E_f(R)=\frac{1}{L}\sum_i \left (E_i - E_{bulk} \right )
\label{eq:ef}
\end{equation}
where the summation is taken over all atoms $i$ contained in a cylinder of
radius $R$ and length $L$ around the dislocation line. $E_{bulk}$ is the
average atomic energy in the random alloy and $E_i$ is the energy of atom
$i$. We now decompose the formation energy into two terms,
$E_f(R)=E_{el}(R)+E_{seg}(R)$, where the elastic and the segregation components
are defined as

\begin{eqnarray}
\label{eq:eel}
E_{el}(R) = \frac{1}{L}\sum_i \left ( E_i - E(c_i,c'_i) \right ),
\\
\label{eq:eseg}
E_{seg}(R)=\frac{1}{L}\sum_i \left ( E(c_i,c'_i)-E_{bulk} \right ).
\end{eqnarray}
Here, we define $E(c_i,c'_i)$ as the energy of a site with average occupancy
$c_i$ while the average occupancy of the neighboring sites is
$c'_i$. $E_{el}(R)$ represents the energy cost of including a dislocation in a
segregated alloy (i.e. an alloy with a compositional fingerprint which is
identical to that produced by a dislocation). $E_{seg}(R)$ is the energy of
segregation per unit length of dislocation (i.e. the energy cost of creating
the compositional fingerprint from a random alloy). For elemental solids or
ordered heteropolar alloys, this is either zero or assumes one of two distinct
values and can thus be taken into account implicitly \cite{blumenau02}. For a
segregated alloy, however, this decomposition is a necessary step in linking
atomistic and continuum concepts.

\begin{figure} 
\begin{center}
\includegraphics[width=0.9\columnwidth]{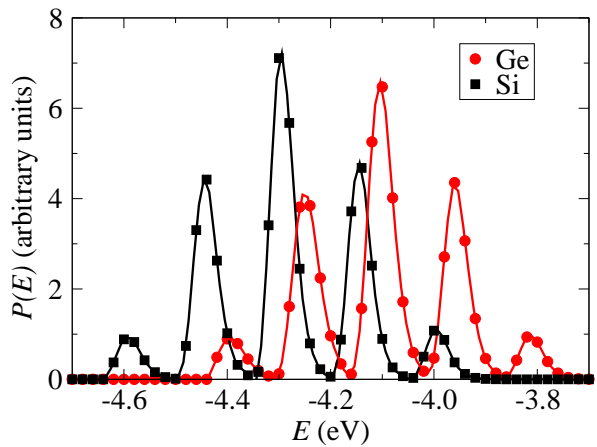} 
\end{center}
\caption{(color online) Probability density of atomic energies for Si
  (squares) and Ge (circles) in a Si$_{0.5}$Ge$_{0.5}$ alloy at 300 K.}
\label{fig:alloy}
\end{figure}

To calculate the elastic and segregation energies defined by
Eqs. (\ref{eq:eel}) and (\ref{eq:eseg}), we first evaluate the function
$E(c,c')$. This is achieved by equilibrating a random Si$_{0.5}$Ge$_{0.5}$
alloy at 300 K using a 4096-atom cubic supercell. All atoms are fourfold
coordinated, and the average cohesive energy was found to be $-4.19$ eV/atom.
In Fig.\ \ref{fig:alloy}, we plot the probability distributions $P(E)$ of atom
energies. For each type of atom (Si or Ge), we observe five distinct and
narrow peaks.  By inspecting the environment of atoms whose atomic energies
are close to these peak values, we conclude that each peak corresponds to a
different chemical environment. For Si, the first peak is at $-4.59$ eV and
corresponds to Si atoms neighboring with four Si atoms; this number coincides
with the calculated cohesive energy of Si at 300 K. The second peak, at
$-4.44$ eV, corresponds to Si atoms neighboring with three Si atoms and one Ge
atom. The third peak, at $-4.29$ eV, corresponds to Si atoms neighboring with
two Si atoms and two Ge atom. The last two peaks correspond to Si atoms
neighboring three Ge atoms (energy $-4.14$ eV) and Si atoms neighboring with
four Ge atoms (energy $-4.00$ eV).  The five peaks in the distribution of atom
energies for Ge atoms can be attributed to the same bonding environments.  The
peak energies and the corresponding composition of the neighbors are: $-4.39$
eV for four Si neighbors, $-4.25$ eV for three Si and one Ge neighbor, $-4.11$
eV for two Si and two Ge neighbors, $-3.95$ eV for three Si and one Ge
neighbor and -3.81 eV for four Ge neighbors. The last number coincides with
the calculated cohesive energy of Ge at 300 K.  This analysis yields values of
the function $E(c,c')$ for $c=$0 or 1 and $c'=$ 0.00, 0.25, 0.50, 0.75 and
1.00; we use linear interpolation to calculate values of $E(c,c')$ for all
other values of $c$ and $c'$.

We can now evaluate $E_{el}(R)$ and $E_{seg}(R)$ as a function of $R$.  The
strain field contributing to the elastic energy term, $E_{el}(R)$, will
involve the superposition of the strain fields from the dislocation and the
compositional segregation evident in Fig. \ref{fig:compos}. This compositional
segregation, consisting of spatially separated regions of compression and
tension, is clearly associated with a strain field of a dipole-like
character. It will consequently fall off relatively quickly compared with the
long-range dislocation strain field. With this in mind, it is appropriate to
define several spatial regimes associated with the segregated dislocation: The
first regime is associated with the dislocation core radius, $R_c$, where
continuum elasticity breaks down. For segregated systems an extended core
region, also exists, where significant segregation is observed.  A third
intermediate regime corresponds to a superposition of the compositional dipole
and dislocation strains. Finally, the far-field region is dominated by the
strain field for the dislocation alone. With the intention of bringing
together continuum and atomistic descriptions we therefore propose a
parameterization of the form
\begin{equation}
E_{el}(R)=b^2 k \ln(R/R_c)+E_c,
\label{eq:er}
\end{equation}
where $R_c$ and $E_c$ are the respective core radius and energy associated
with a dislocation of Burgers vector ${\bf b}$, and $k$ should depend on the
region of study. In the case of an ideal continuum model of an elemental
crystal, compound or ordered alloy, the factor $k$ is formally related to
elastic constants of the crystal and geometry of the dislocation
\cite{hirthbook}. In the case of segregated systems, $k$ is expected to assume
the random alloy value in the far-field region. In the intermediate and
extended core regimes, the dipole segregated region around the dislocation
will modify the energetics and $k$ must be considered as an effective
parameter which must be determined from atomistic calculations. Since the size
of our simulation cell seems to be limited to the intermediate regime, the values of $k$ we calculate should be
smaller than the value of $k$ for the non-segregated alloy.

\begin{figure} 
\begin{center}
\includegraphics[width=0.9\columnwidth]{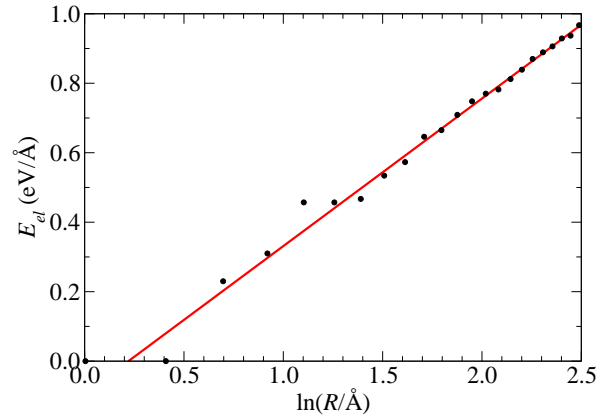} 
\end{center}
\caption{(color online) Elastic energy, $E(R)$, contained in a cylinder of
  radius $R$ and unit length around the dislocation core versus the logarithm
  of $R$ for a 60$^\circ$ glide dislocation in a Si$_{0.5}$Ge$_{0.5}$ alloy at
  300 K. }
\label{fig:e_vs_lnr}
\end{figure}

In Fig.\ \ref{fig:e_vs_lnr} we plot $E_{el}(R)$, as calculated using
Eq. (\ref{eq:eel}) as a function of ln$R$ in the intermediate regime. We find
that the thus defined $E_{el}(R)$ is indeed a smooth, linear function of
ln$R$, consistent with the parameterization scheme of Eq. (\ref{eq:er}) which
effectively takes into account the influence of the extended core. We can then
determine the core radius as the distance at which the linearity of
$E_{el}(R)$ breaks down. In Fig.\ \ref{fig:e_vs_lnr}, for example, this occurs
at ln($R$/\AA)=1.5.  The core energy is then defined as $E_c=E_{el}(R_c)$,
while $k$ is determined by the slope of the linear region, in accordance with
Eq. (\ref{eq:er}). The error in the somewhat subjective determination of
ln($R_c$/\AA) is about 0.05, giving an error bar of 0.05 eV/\AA\ for $E_c$.
By taking the average over many runs, we determine dislocation core parameters
for the alloy to be $k$ = 4.5 GPa, $R_c$= 4.5 \AA\ and $E_c=E_{el}(R_c)$=0.59
eV/\AA. As the core parameters describe local deformations around the
dislocation line, they are not very sensitive to a macroscopic strain field:
For the (100)-constrained alloy, we find a marginaly lower core energy (0.57
eV/\AA) and core radius (4.2 \AA). Temperature also seems to play a minor role
in the core energies: at 900 K, the core energy is only 0.03 eV/A higher than
that at 300 K. 

We have evaluated the radial dependence of $E_{seg}(R)$ directly from
Eq. (\ref{eq:eseg}) and find a relatively weak dependence on $R$ which can be
understood from Figs. \ref{fig:compos} and \ref{fig:alloy}. Terms in
Eq. (\ref{eq:eseg}) will be of opposite sign depending on whether $E(c_i,
c'_i)$ is associated with Ge-rich or Si-rich regions. Consequently, they will
tend to cancel in the summation over a cylindrical region about the
dislocation line, reflecting the symmetry of the segregated regions (Fig.\
\ref{fig:compos}). Therefore, we find that $E_{seg}(R)$ can be approximated by
a constant value of the order 0.2 eV/\AA. Adding this value to
Eq. (\ref{eq:er}) shows that the dislocation formation energy in segregated
alloys can be approximated by the continuum energy expression for a
dislocation in an elemental crystal with a judicious choice of parameters
appropriate to specific spatial regimes of the segregated dislocation. This
surprising result indicates that simple continuum concepts, including core
energies and radii, are still of value in situations where alloy segregation
is significant and that the effects of the compositional dipole strain can be
included specifically in $k$.

We now apply the above framework for determining core parameters in segregated
systems to provide additional insight into segregation energetics by direct
comparison with dislocations in idealized random alloys. This is achieved by
performing simulations for dislocations in Si$_{0.5}$Ge$_{0.5}$ alloys
artificially constrained to have a random occupation of the lattice sites. We
find that the dislocation formation energy (3.7 eV/\AA) is 2.0 eV/\AA\ greater
than in the segregated alloy (1.7 eV/\AA). Surprisingly, however, the ideal
random alloy has a smaller core radius of 3.9 \AA\ with a correspondingly
lower core energy of 0.31 eV/\AA, as determined by a linear fit of
Eq. (\ref{eq:er}). The parameter $k$ of Eq. (\ref{eq:er}) is found to be 5.4
GPa. This value is, larger than the value found for the segregated alloy, as
expected.

To understand this we note that preventing atoms to exchange identities and
occupy their energetically favored sites around the dislocation core results
in weaker bonds in this area which are easier to deform. This explains the
lower core energies associated with perfectly random alloy systems.  However,
despite higher core energies, the significant segregation outside of the core
radius $R_c$ (Fig.\ \ref{fig:compos}) lowers the overall dislocation
formation energy for segregated systems by reducing the overall elastic energy
$E_{el}(R)$.
 
In summary, by partitioning the dislocation formation energy into elastic and
segregation components we have established a general framework for calculating
the core properties of dislocations in segregated alloys. This approach opens
up new possibilities for the study of more complex processes such as `alloy
drag' on the moion of dislocations \cite{wang00}, the influence of local
segregation (including vacancies) on the Peierls barrier and kink mobility
\cite{farkas}, as well as dislocation-induced segregation in confined systems
such as quantum dots.

I. R. acknowledges useful discussions with Dr. J. Schiotz. We are particularly
grateful to an anonymous referee for providing valuable insight into the
continuum description of segregated dislocations. This work is supported by a
grant from the EU and the Ministry of National Education and Religious Affairs
of Greece through the action ``$\mathrm{E\Pi EAEK}$'' (programme
``$\mathrm{\Pi Y\Theta A\Gamma OPA\Sigma}$.'') D. E. J.  is funded by the ARC.


\begin{thebibliography}{10}

\bibitem{hull89}
R. Hull, J.~C. Bean, and C. Buescher, J. Appl. Phys. {\bf 66},  5837  (1989).

\bibitem{hirthbook}
J.~P. Hirth and J. Lothe, {\em Theory of Dislocations} (McGraw Hill, 1968).

\bibitem{arias94}
T.~A. Arias and J.~D. Joannopoulos, Phys. Rev. Lett. {\bf 73},  680  (1994).

\bibitem{blase00}
X. Blase {\it et~al.}, Phys. Rev. Lett. {\bf 84},  5780  (2000).

\bibitem{blumenau02}
A.~T. Blumenau {\it et~al.}, Phys. Rev. B {\bf 65},  205205  (2002);
A.~T. Blumenau {\it et~al.}, J. Phys: Cond. Matt. {\bf 14},  12741  (2002).

\bibitem{martinelli04}
L. Martinelli {\it et~al.}, Appl. Phys. Lett. {\bf 84},  2895  (2004).

\bibitem{houghton90} See, for example, D. C. Houghton {\it et~al.},
J. Appl. Phys. {\bf 67} 1850 (1990).

\bibitem{ovidko02}
I.~A. Ovid'ko, Phys. Rev. Lett. {\bf 88},  046103  (2002).

\bibitem{denker05} 
U. Denker and D. E. Jesson, Phys. Stat. Sol. (b) {\bf 242}, 2455 (2005).

\bibitem{tersoff89}
J. Tersoff, Phys. Rev. B {\bf 39},  5566  (1989).

\bibitem{kelires89}
P.~C. Kelires and J. Tersoff, Phys. Rev. Lett. {\bf 63},  1164  (1989).

\bibitem{kelires95-98}
P.~C. Kelires, Phys. Rev. Lett. {\bf 75},  1114  (1995); Surf. Sci. {\bf 418},
  L62 (1998).

\bibitem{marzegalli05}
A. Marzegalli, F. Montalenti, and L. Miglio, Appl. Phys. Lett. {\bf 86},
  041912  (2005).

\bibitem{ASE} The dislocation strain field is introduced by employing the
  Atomic Simulation Environment, ASE; see {\tt
  http://www.camp.dtu.dk/Software}.

\bibitem{wang00} Y. Wang {\it et~al.}, Acta mater. {\bf 48}, 2163 (2000).

\bibitem{farkas} J. K. Ternes, D. Farkas and R. Kriz, Phil.  Mag. A {\bf 72},
  1671 (1995); D. Farkas {\it et~al.}, Acta Materialia {\bf 44}, 409 (1996).
 
\end{thebibliography}
\end{document}